\documentstyle[11pt,dina4,twoside]{article}
\newcommand{\refkl}[1]{(\ref{#1})}

\renewcommand{\v}[1]{\underline{#1}} 
\newcommand{\m}[1]{\underline{\underline{#1}}}           

\newcommand{\nab}{\v{\nabla}}

\renewcommand{\d}[1]{\partial_{#1}}                     

\newcommand{\vonrt} {\mbox{$(\v{r},t)$}}                

\newcommand{\barsig}{\overline{\sigma}}
\newcommand{\barphi}{\overline{\phi}}
\newcommand{\barn}{\overline{n}}
\newcommand{\dEo}{\delta E_0}
\newcommand{\Dg}{d(\gamma)}

\newcommand{\drhoo}{\delta \rho_0}

\newcommand{\dsigo}{\delta\sigma_0}
\newcommand{\epsa}{\epsilon_{a}}
\newcommand{\epsq}{\epsilon_{q}}
\newcommand{\epsp}{\epsilon_{\perp}}
\newcommand{\epso}{\epsilon_{0}}

\newcommand{\nhocho}{n^{(0)}}
\newcommand{\mup}{\mu^{+}}
\newcommand{\mum}{\mu^{-}}
\newcommand{\np}{n^{+}}
\newcommand{\nm}{n^{-}}
\newcommand{\opsig}{\hat{\sigma}}
\newcommand{\opeps}{\hat{\epsilon}}
\newcommand{\opsigq}{\hat{\sigma}_q}
\newcommand{\opepsq}{\hat{\epsilon}_q}
\newcommand{\opKq}{\hat{K}_q}
\newcommand{\phihochi}{\phi^{(1)}}
\newcommand{\phihochmi}{\phi^{(-1)}}
\newcommand{\Sgi}{s_1(\gamma)}
\newcommand{\Sgii}{s_2(\gamma)}
\newcommand{\siga}{\sigma_{a}}
\newcommand{\sigq}{\sigma_{q}}
\newcommand{\sigo}{\sigma_{0}}
\newcommand{\sighocho}{\sigma^{(0)}}
\newcommand{\sigp}{\sigma_{\perp}}
\newcommand{\talpha}{\tilde{\alpha}}
\newcommand{\tr}{\tilde{r}}
\newcommand{\taudo}{\tau_d^{(0)}}
\newcommand{\tildesig}{\tilde{\sigma}}
\newcommand{\tilderho}{\tilde{\rho}}
\newcommand{\Vco}{V_c^{(0)}}

\newcommand{\sectionILCC}[1]{\noindent \underline{#1}\vspace{2mm}\\}

\begin{document}

\renewcommand{\baselinestretch}{1.5} \small\normalsize

\hspace{10mm}
  \parbox{130mm}{
BIPOLAR ELECTRODIFFUSION MODEL FOR \vspace{-3mm}\\
ELECTROCONVECTION IN NEMATICS
}

\vspace{15mm}

\hspace{10mm}
  \parbox{130mm}{
MARTIN TREIBER AND LORENZ KRAMER \vspace{-3mm}\\
Physikalisches Institut, Universit\"at Bayreuth,
95440 Bayreuth, Germany
}

\vspace{15mm}

\hspace{10mm}
\begin{minipage}{130mm}
\renewcommand{\baselinestretch}{1}
\small \normalsize  
\underline{Abstract}
\ \ The common description of the electrical behavior of a nematic
liquid crystal as an anisotropic dielectric medium  with (weak) ohmic
conductivity
is extended to an electrodiffusion model with two active ionic species.
Under appropriate, but rather general conditions the additional effects
can lead to a distinctive change of the threshold behavior of the
electrohydrodynamic instability, namely to travelling patterns instead of
static ones. This may explain the experimentally observed phenomena.
\end{minipage}

\vspace{15mm}


\sectionILCC{1. INTRODUCTION}
In the last years electrohydrodynamic convection (EHC) in nematic
liquid crystals (NLC) has become a standard physical phenomenon for
investigating
pattern formation and the paradigma for anisotropic
systems \cite{kramer_jphys,zimmermann,rehberg_fk,pesch}.
The main advantage of EHC compared to other pattern-forming systems
are the short time scales, the possibility to make cells with very
large aspect ratios and two easily adjustable control parameters
(AC voltage $V_{eff}$
and frequency $\omega_{ext}$), leading to a wealth of instability scenarios
\cite{rehberg_fk,pesch}.
For a planar (homogeneous) alignment of the director at the
confining plates (electrodes), one observes at the instability from the
unstructured state (primary instability)
normal or oblique rolls, i.e. roll axis
normal or tilted with respect to the equilibrium director \cite{kramer_jphys}.
Both patterns can be stationary
or travelling \cite{kai,rehberg_travV,joets,rehberg_travMBBA,dennin};
furthermore the bifurcation can be continuous or (very weakly) hysteretic
\cite{rehberg_fluctPRL}.

The scenario appearing above threshold depends on $\omega_{ext}$,
the thickness $d$ of the cell, the material
and the temperature. Especially the conductivity (a material
parameter easily adjustable by doping),
the dielectric anisotropy and, as will be shown below,
other electrical properties of the material are relevant.

The commonly used theory to describe EHC are the
Ericksen-Leslie equations \cite{erickson,leslie} combined with the
quasi-static Maxwell equations, thereafter referred to as the standard model
\cite{kramer_jphys}.
This theory describes quantitatively (in many cases) the
stationary features like the threshold voltage, the wavenumber
of the rolls and the transition from normal to oblique rolls
in particular for sufficiently low
external frequencies (conductive regime).
The main discrepancy between the predictions of the standard model
and the experiments, however, is the observed Hopf bifurcation.
The linear analysis
leads for a planar configuration always
to a continuous bifurcation to stationary rolls \cite{kramer_jphys}.
It is even difficult to
produce with this model a primary Hopf bifurcation by changing rather
drastically some material parameters \cite{zimmermann}.
The discrepancies are pronounced for thin cells, pure materials
(low conductivity) and high external frequencies.

In this work we replace in the electrical part of the standard model
the ohmic conductivity
by two dynamically active species of charge carriers
(one positive, one negative). The constant ohmic conductivity
of the standard model becomes a dynamical variable on its own
with Ohm's law replaced by migration and diffusion
parts of the current. Together with the charge density $\rho$ (coupled to
the potential of the induced field via the Poisson equation) the new
model contains two electrical variables.
The new physical processes incorporated in the theory are ionic migration,
diffusion and dissociation-recombination. The corresponding new material
parameters are mobilities, diffusion constants and a recombination rate
constant.
The associated time and length scales are, respectively, the migration time
$\tau_{mig}$ needed for an ion to
travel through the cell, the (Debye) diffusion length
$\lambda_D$ and the (linear) recombination time $\tau_{rec}$.
If $\tau_{rec}$ is much smaller than the other time scales the conductivity
can be eliminated and one has $\rho$ as the only electrical dynamic variable.
If in addition $\tau_{mig}$ is large and $\lambda_D$ is small, drift and
diffusion can be neglected and we are back to ohmic conductivity and
the standard model.
Comparing  $\tau_{rec} \propto 1/n_0$ ($n_0 =$ equilibrium density of charge
carriers of each species) and
$\tau_{mig} \propto d^2/V_{eff}$ with the director
relaxation time $\tau_d \propto d^2$, and $\lambda_D \propto n_0^{-1/2}$
with the roll dimensions,
it is evident, that
our model differs from the standard model just in the parameter ranges
where the latter fails to describe the experiments.

In Sec. 2 we present the basic equations and assumptions. In Sec. 3
the model is applied to the basic state and linearized around this state.
Secion 4 contains an approximate calculation showing a Hopf bifurcation
for sufficiently thin cells, clean materials or sufficiently
high external frequencies (for materials with negative dielectric anisotropy).
Sec. 5 gives a discussion and a brief comparison with experimental results.
\vspace{8mm}
\clearpage

\sectionILCC{2. BASIC EQUATIONS}
It seems reasonable to assume that the conductivity of  NLC's is
caused by two species of ionic charge carriers arising from dissociation
of impurities or of the NLC itself \cite{richardson}. We assume
charges $\pm e$ (no loss of generality) and overall neutrality
$\int \! d^3\v{r}\np=\int \! d^3\v{r}\nm$ with the number densities
$\np$ and $\nm$.
The electrical current $\v{J}^{\pm}$
of each species is caused by
migration in the electrical fields $\v{E}$,
diffusion due to carrier-density gradients, and advection with the fluid
velocity $\v{v}$,
$\v{J}=\v{J}^+
+\v{J}^-$ with $
\v{J}^{\pm}=
\v{J}^{\pm}_{mig}+
\v{J}^{\pm}_{diff}+
\v{J}^{\pm}_{adv}$
\cite{turnbull}.
In uniaxial-anisotropic systems we have
\begin{eqnarray}
\label{jmig}
  \v{J}_{mig}&=& e (\m{\mu}^+\np+\m{\mu}^-\nm)\v{E} =: \m{\sigma}\ \v{E}, \\
\label{jdiff}
  \v{J}_{diff} &=& -e (\m{D}^+ \nab \np - \m{D}^- \nab \nm), \\
\label{jconv}
  \v{J}_{adv}&=&e  (\np - \nm) \v{v} =: \rho\v{v},
\end{eqnarray}
where $\rho$ is the charge density and $\m{\sigma}$ the tensor of the
local conductivity.
The mobility tensors are
$ \mu_{ij}^{\pm}=
\mu_{\perp}^{\pm}\delta_{ij} +
\mu_{a}^{\pm}n_in_j$
($\v{n}$ is the director field) with $\mu_{\perp}^{\pm}, \mu_{a}^{\pm}$
essentially independent of $\np, \nm$ and $\v{E}$
\cite{richardson}. We assume that the relative anisotropies for the two
ionic species are the same and thus equal to the relative
conductivity anisotropy, $\sigma_a/\sigma_{\perp}$, as measured
in the low-field ohmic range,
\begin{equation}\label{mu}
\mu_{ij}^{\pm}=\mu_{\perp}^{\pm} (\delta_{ij}+
\frac{\sigma_a}{\sigma_{\perp}}n_in_j)
 =: \mu_{\perp}^{\pm} \sigma'_{ij}.
\end{equation}
We adopt the Einstein law
\begin{equation}
\label{einstein}
\m{D}^{\pm} = V_T\m{\mu}^{\pm},
\end{equation}
where $V_T=k_BT/e \approx 26 \mbox{mV}$ is the thermal potential.
Experimentally this relation is satisfied within a factor of about 3
\cite{naito_1,naito_2,naito_buda}.

The dissociation-recombination process AB $\rightleftharpoons \mbox{A}^+ +
\mbox{B}^-$ with the undissociated molecules or impurities AB and the ions
$\mbox{A}^+,\mbox{B}^-$ \cite{richardson} is described
in homogeneous systems (no spatial variation)
by the rate equations
$\dot{n}^{\pm}=k_d n_{AB}-k_r \np\nm$ ($k_d, k_r=$dissociation and
recombination constants) and $\dot{n}_{AB}=-\dot{n}^{\pm}$.
In the weak-dissociation limit $\np,\ \nm << n_{AB}$ the concentration
$n_{AB}$ of the undissociated molecules
is nearly constant in time, and can be replaced by
its equilibrium value which we write as
$K_{diss} n_{AB}^{eq}=n_0^2$, where $K_{diss}=k_d/k_r$ is the dissociation
constant and $n_0$ is the equilibrium value of $\np$ and $\nm$.
This assumption is usually valid when $K_{diss} <<n_0$.
For MBBA with ionic dopant TBATPB Ref. \cite{richardson} gives
$K_{diss} =3.4\times 10^{21}m^{-3}$ while equilibrium densities
$n_0=\sigma_{eq}/(e(\mup + \mum))$ are of the order of $10^{19}m^{-3}$
for typical mobility and conductivity values (Tables I and II).

\vspace{8mm}
\begin{tabular}{ll|c|c}
\multicolumn{4}{l}{TABLE I Scaling} \\ \hline
\hspace{10mm} &        
  quantity         & scaling unit              & typical value $^{+}$  \\
\hline
& lengths          & $d/\pi$                   &
       $ \pi^{-1}(10 ... 100 \mu$m)     \\
& time             & $\tau_d^{(0)} = \frac{\gamma_1 d^2}{K_{eff} \pi^2}$
                        & 0.073 s $\left( \frac{d}{10\mu m}\right)^2$  \\
& voltage          & $\Vco=\sqrt{\frac{\pi^2 K_{eff}}{\sigma_a\tau_q}}$
                                               & 2.7 V                \\
& conductivities   & $\sigma_{eq}=\mu e n_0$   &
      $10^{-7} ... 10^{-9} (\Omega \mbox{m})^{-1}$                    \\
& dependent quantities $E $ and $\rho $
    & $\frac{\Vco\pi}{d}$ and $\frac{\Vco\epsilon_0\epsp \pi^2}{d^2}$ & \\
& orientational elasticities & $K_{eff}=K_{11}+K_{33}$ & $15.3\times 10^{-12}$N
\\
& dielectric constants       & $\epsilon_0\epsilon_{\perp}$
                 & $4.2\times 10^{-11}\mbox{As}/\mbox{Vm}$\\
& viscosities       & $\gamma_1=\alpha_3-\alpha_2$
                                             & 0.11 kg m s$^{-1}$ \\ \hline
& \multicolumn{3}{l}{\parbox{100mm}{$^{+}$ parameter set MBBA I
                                    \cite{kramer_jphys}
                                    }
}    
\\ \hline
\end{tabular}

\vspace{8mm}
\begin{tabular}{ll|c|c}
\multicolumn{4}{l}{TABLE II material parameters related to conduction} \\
\hline
\hspace{10mm} &        
parameter &
       material    &  value and source \\ \hline
& mobility $\mu$   $ ^{\ast}$
     & MBBA        &  0.371 \cite{richardson}; \ 1 \cite{turnbull}  \\
&    & 5CB         &  0.06...0.35 \cite{naito_2,naito_3,naito_buda}  \\
& mobility ratio $\gamma$
     & $\mbox{MBBA}^{**}$  & 1\ \cite{richardson}                            \\
&    & 5CB         & $<<1$  \ \cite{naito_buda}                 \\
& mobility anisotropy $\frac{\mu_a}{\mu_{\perp}} = \frac{\siga}{\sigp}$
     & MBBA        & 0.33 \ \cite{richardson}; \ 0.5 \ \cite{sigasource}\\
& recombination-
     & 5CB         &$10\mbox{s}^{-1}n_0^{-1}$\ \cite{naito_3}              \\
& rate constant $k_r$
     & dielectric  & $10^{15}\mbox{m}^{3}\mbox{s}^{-1}$ \ \cite{turnbull};
                                                       \vspace{-2mm} \\
&    & liquids     &                    \\
& dissociation constant $K_{diss}$
     & $\mbox{MBBA}^{**}$ & $ 3.4 \times 10^{21}\mbox{m}^{-3}$
        \cite{richardson}                                  \\ \hline
& \multicolumn{3}{l}{\parbox{100mm}{$^{\ast}$ in units of
  $10^{-9}m^2/(Vs)$ \hspace{10mm} $^{**}$ Doping TBATPB.}
}    
\\ \hline
\end{tabular}
\vspace{10mm}

In the spatially extended case the rate equations are generalized to
\begin{equation} \label{space}
\d{t}n^{\pm} \pm \frac{1}{e}\nab\cdot \v{J}^{\pm} = \dot{n}^{\pm},
\end{equation}
or with (\ref{jmig}) - (\ref{einstein}) and $n_{AB}=n_{AB}^{eq}$
\begin{equation}\label{nbalance}
\d{t}n^{\pm} + \nab\cdot\left[
  \v{v}n^{\pm}+ \mu_{\perp}^{\pm}\m{\sigma}'(\pm \v{E}-V_T\nab)n^{\pm}
  \right]
  = k_r(n_0^2-\np\nm).
\end{equation}
In view of coupling Eqs.\refkl{nbalance} to the director and momentum-balance
equations it is convenient to write Eqs.\refkl{nbalance} as a
continuity equation for the charge
density $\rho=e(\np-\nm)$ and a balance  equation for the conductivity
$\sigma:=\sigma_{\perp} = e(\mu_{\perp}^+\np + \mu_{\perp}^-\nm)$,
\begin{eqnarray}
\label{bipol_rho}
  \d{t}\rho & + &  \nab\cdot\left[
  \rho\v{v} + \m{\sigma}'\v{E}\sigma -
  V_T \m{\sigma}'\nab
    \left(\Dg\sigma+2\mu\Sgi\rho
    \right)
  \right] = 0, \\
\label{bipol_sig}
  \d{t}\sigma & + &
  \nab\cdot\left[
  \sigma\v{v}+\mu\m{\sigma}'\v{E}(\Dg\sigma+\mu\Sgi\rho) -
  \mu V_T \m{\sigma}'\nab\left(
    \Sgii \sigma + \Dg \mu\Sgi\rho
    \right) \right] \nonumber \\
&=&
  k_r n_0\sigma_{eq}
    \left[1-\frac{(\sigma+\mu_-\rho)(\sigma-\mu_+\rho)}{\sigma_{eq}^2}
    \right],
\end{eqnarray}
where we introduced the effective mobility $\mu=\mu_{\perp}^+ + \mu_{\perp}^-$,
the mobility ratio $\gamma=\mu_{\perp}^-/\mu_{\perp}^+$ together with
$\Dg=(1-\gamma)/(1+\gamma)$, $\Sgi=\gamma/(1+\gamma)^{2}$ and
$\Sgii=(1+\gamma^2)/(1+\gamma)^2$,
and the equilibrium conductivity $\sigma_{eq}=\mu e n_0$.
With the Poisson equation
\begin{equation}\label{poisson}
\rho=\nab\cdot(\epsilon_0\epsilon_{\perp}\m{\epsilon}'\v{E}),
  \ \ \m{\epsilon}'=\delta_{ij}+\frac{\epsilon_a}{\epsilon_{\perp}}n_in_j
\end{equation}
and the potential for $\v{E}$ (see Sec. 3)
all electrical variables are given in terms of
the (scalar) potential and $\sigma$.

We need five boundary conditions (BCs) for the electrical variables in
addition to the usual no-slip planar BCs
$\v{n}=(1,0,0), \v{v}=0$ at the confining plates
($z=\pm d/2$), which act as electrodes (NLC layer of infinite extent
in the $x$-$y$ plane) .
The applied voltage is given by
\begin{equation}\label{bc_voltage}
\int^{d/2}_{-d/2}dz \ E_z = V(t).
\end{equation}
Furthermore there are relations between current, electrical
field and density for each species at the electrodes which can depend
in a complicated way on electrochemical processes and may be parametrised
e.g. for $z=d/2$ as
$J_z^{\pm}=\sigma^{\pm}_{surface}E_z - D^{\pm}_{surface}
(n^{\pm}_{ext}-n^{\pm})$.
Two limiting cases are i) strongly injecting electrodes, i.e. at
least one $\sigma_{surface}$ is very large,
which leads to $E_z=0$ (space-charge limiting conditions) and is
(in the isotropic-unipolar case) adopted e.g. in
Refs \cite{turnbull,felici} and
ii) no transfer of any charge through the
electrodes (blocking electrodes),
\begin{equation}\label{bc_current}
J_z^{+}(z=\pm d/2)=
J_z^{-}(z=\pm d/2)=0.
\end{equation}
These last BCs do not involve unknown electrochemical processes
and will be assumed in the rest of this paper.
Blocking electrodes imply that
the averaged charge density per area,
$Q:= lim_{A\to \infty} \frac{1}{A}\int_A dx\ dy \int_{-d/2}^{d/2}dz \rho$
or, equivalently, $E(d/2)-E(-d/2)$ is constant and we can
incorporate permanently adsorbed charges (for future purpose) by setting
$Q=Q_{ad}+\int dz \rho=0$ or with the Poisson equation (\ref{poisson})
\begin{equation}\label{adsorb}
E_z(d/2)-E_z(-d/2)= - \frac{Q_{ad}}{\epsilon_0\epsilon_{\perp}}.
\end{equation}
Equations (\ref{bipol_rho})and (\ref{bipol_sig}) together with
the director and momentum balance, and incompressibility
equations of the
standard model \cite{kramer_jphys},
the boundary conditions (\ref{bc_voltage}),(\ref{bc_current})
and the charge adsorption (\ref{adsorb}), determining the conserved quantity
$E_z(d/2)-E_z(-d/2)$ represent
the bipolar electrodiffusion model. It contains in addition to $\sigma_{\perp}$
and
$\sigma_a$ three more conduction parameters $\mu_{\perp}^+, \gamma$ and $k_r$
of the NLC at a given temperature
(see Table II).
Equations \refkl{bipol_rho} and \refkl{bipol_sig} are coupled
to the fluid velocity
by the advection term and to the director by $\m{\sigma}'$ and
$\m{\epsilon}'$. Conversely, the director and momentum-balance
equations of the standard model
are coupled to the electrical fields via the electrical torque $\propto
\epsilon_0\epsilon_a(\v{n}\cdot\v{E})\v{E}$ on the director
and the electrical
volume force $\nab \ \m{T}^{el} = \rho \v{E}+h.o.t.$ on the fluid.
The director and momentum-balance equations do not couple directly to $\sigma$.

To see the relative magnitude and time scales of the different
processes we scale all variables and parameters according to Table I.
The result is
\begin{eqnarray}
\label{scal_rho}
  P_1(\d{t}+ \v{v}\cdot\nab) \rho
  & = &
   -  \nab\cdot\m{\sigma}'\v{E}\sigma
   + D\nab\cdot\m{\sigma}'\nab
             \left[\alpha^{-1}\Dg\sigma +2\Sgi\rho \right],
    \\
\label{scal_sig}
  P_1(\d{t}+ \v{v}\cdot\nab) \sigma
  & = &
   - \alpha \nab\cdot\m{\sigma}'\v{E}
              \left[\Dg\sigma+\alpha\Sgi\rho \right]
   + D\nab\cdot\m{\sigma}'\nab
             \left[\Sgii \sigma
                +  \Dg \alpha\Sgi\rho \right]
      \nonumber \\
&-&
    r\left[
      \left(\sigma + \frac{\gamma\alpha}{1+\gamma}\rho \right)
      \left(\sigma - \frac{\alpha}{1+\gamma}\rho \right)
      - 1 \right], \\
\label{scal_poisson}
  \nab\cdot\m{\epsilon}'\v{E} & = & \rho,
\end{eqnarray}
and scaled director and momentum-balance equations
(eqs.(52) in \cite{treiber_fluct} with a slightly different scaling).
The BCs for an applied voltage $V(t)=\sqrt{2}V_{eff}\cos\omega_{ext}t$
and blocking electrodes are
\begin{equation}\label{scal_bc}
\begin{array}{rcl}
\int_{-\pi/2}^{\pi/2}dz \ E_z&=& V'(t), \\
\left[
  E_z\sigma -
  D\d{z}(\alpha^{-1}\Dg\sigma + 2\Sgi\rho)
  \right]_{z=\pm1/2} &=& 0, \\
\left[
  D\d{z} \sigma -
  \alpha E_z (\Dg\sigma + 2\alpha\Sgi\rho)
  \right]_{z=\pm1/2} &=& 0,
\end{array}
\end{equation}
where the scaled voltage is $V'(t)=V(t)/\Vco = \sqrt{2R}\cos\omega t$, and
we wrote (\ref{bc_current}) as $J_z^+ \pm J_z^- =0$.

Apart from the two external control parameters $R$ and $\omega_0$ the model
contains four system parameters $P_1,\alpha,r$ and $D$, see Table III.
The first three are (Prandtl-number like) time-scale ratios relating
the time scales of director relaxation,
charge relaxation, carrier-generation-recombination kinetics and
carrier migration, and $D$ is (up to a factor $\sqrt{2}$) the diffusion length
(Debye screening length) $\lambda_D=(\mu V_T\tau_q/2)^{1/2}$
in units of $d/\pi$. The diffusion is not an independent parameter since
it is coupled to $\alpha$ via the scaled Einstein law $D=\alpha V_T/\Vco$.


\begin{tabular}{lc|c|c|c}
\multicolumn{5}{l}{TABLE III System parameters of the bipolar model} \\ \hline
& parameter        &   physical process   &\ typical value$^{\#}$ \
                                          &\  proportionality        \\ \hline
& $\sqrt{R}=\frac{V_{eff}}{\Vco}$  & control parameter   &$\geq 2.5$ &   \\
& $\omega_0 = \omega_{ext}\taudo$ & control parameter & $1 ... P_1^{-1}$
                                &                                    \\
& $P_1 = \frac{\tau_q}{\taudo} = \frac{\epsilon_0\epsp}{\sigp\taudo}$
                                & charge relaxation   & 0.057
                                & $(\sigma_{\perp}d^2)^{-1}$          \\
& $\alpha = \frac{\tau_q}{\tau_{mig}(\Vco)}
    =\mu \Vco\tau_q\frac{\pi^2}{d^2}$
                                & ion migration        & 1.1
   & $\frac{\Vco}{n_0d^2} = \frac{\mu e \Vco}{\sigma_{\perp}d^2}$
                                          		               \\
& $r= \frac{\tau_q}{2\tau_{rec}}=k_r n_0 \tau_q$
                                &recombination         &   $\le 1$
                 &$\frac{k_r n_0}{\sigma_{\perp}}=\frac{k_r}{\mu e}$     \\
& $\sqrt{D}=\sqrt{2}\pi \frac{\lambda_D}{d} =
  \frac{\pi}{d}\sqrt{\frac{V_T\epsilon_0\epsp}{e n_0}}$
                                & diffusion            & $0.1$
                             & $(n_0d^2)^{-1/2} \propto \alpha^{1/2}$ \\ \hline
& \multicolumn{4}{l}{\parbox{120mm}{$^{\#}$ parameter set MBBA I with
  \vspace{-3mm}\\
   $\sigma_{\perp}=10^{-8}(\Omega\mbox{m})^{-1}, \mu=10^{-9}m^2/(Vs)$,
    $d=10 \mu $m, and $\gamma=1$.}
}    
\\ \hline
\end{tabular}
\vspace{10mm}

The standard theory is recovered when $\sigma - \sigma_{eq}$ remains negligibly
small (and diffusion can be neglected). This is the case when
i) the recombination time
$\tau_{rec}=(2 k_r n_0)^{-1}$ is much smaller than the
charge relaxation time and ii) the ion migration time to travel
through the cell at the applied voltage is much larger than the
charge relaxation time and the time scale of the external frequency.
This essentially means $r>>1$ and $\alpha<<1$, see Table III.
\vspace{8mm}

\sectionILCC{3. BASIC STATE AND LINEAR ANALYSIS}
In contrast to the standard model the electrodiffusion model actually implies
a nontrivial basic state $\sigma_0(z,t), \ \drhoo(z,t), \
\v{E}_0=(0,0,E_0(z,t)), \ \v{n}_0=(1,0,0)$ and $\v{v}_0=0$.
For blocking electrodes this is obvious because charge conservation
together with finite conductivity leads to an accumulation
of charges near the electrodes. Other BCs lead to a nontrivial
basic state (for the electrical variables) as well, except ohmic
BCs, $J_z=\sigp E_z$, implicitely assumed in the standard model.

We decompose the fields of the scaled electrodiffusion model into
(i) the basic state of the standard model, (ii) the difference between the
actual
basic state and the standard-model basic state (denoted with a $\delta$),
and (iii) linearized fluctuations,
\begin{equation}
\label{decomposition}
\begin{array}{lll}
  \multicolumn{3}{l}{\v{n}  = (1,0,0) + (0,n_y,n_z) + h.o.t.,} \\
  \sigma & = \sigma_0(z,t) + \tildesig\vonrt &
           = 1 + \delta \sigma_0(z,t) + \tildesig\vonrt, \\
  \v{E}  & = E_0(z,t)\v{e}_z - \nab \phi\vonrt
         & = (\pi^{-1} V'(t) + \delta\v{E}_0(z,t))\v{e}_z - \nab \phi\vonrt, \\
  \rho   & = \delta\rho_0(z,t) + \tilderho\vonrt
         & = \delta\rho_0(z,t) + \opeps \phi + \epsa E_0 \d{x}n_z + h.o.t.,
  \end{array}
\end{equation}
where $\delta\rho_0 = \d{z}\delta E_0$, and
we used \refkl{scal_poisson}, the second equation of \refkl{poisson} and the
definition
$\opeps= -\epsilon'_{ij}(\v{n}_0)\d{i}\d{j}=-(\nab^2+\epsa\d{x}^2)$
to express $\tilde{\rho}$ in terms of $n_z$ and the induced potential $\phi$.

Inserting the basic state into
\refkl{scal_rho}-\refkl{scal_poisson} and using $\d{z}E_0=\delta\rho_0$ gives
the basic-state equations in terms of $E_0$ and the deviations from
the trivial basic state, $\delta \sigma_0, \delta \rho_0$,
\begin{eqnarray}
\label{nontriv_sig}
P_1\d{t}\drhoo &=&
  \left[ -1+2D \Sgi\d{z}^2\right]\drhoo
 +\left[ -E_0\d{z}+\alpha^{-1} D \Dg\d{z}^2\right]\delta\sigma_0
 - \drhoo\dsigo, \\
\label{nontriv_rho}
P_1\d{t}\dsigo &=&
  \alpha\left[(r-1)\Dg - \alpha\Sgi E_0\d{z}
                       + D\Dg\alpha\Sgi\d{z}^2 \right]\drhoo \nonumber \\
 & - &     \left[2 r +\alpha\Dg E_0 \d{z} - D\Sgii\d{z}^2\right]\dsigo
                                                        \nonumber \\
 & - & r\dsigo^2 +\alpha (r-1)\Dg\dsigo\drhoo +\alpha^2(r-1)\Sgi\drhoo^2.
\end{eqnarray}
In regions where $\delta\rho_0 << 1$ and $\delta\sigma_0 << 1$
the volume charge relaxes with the rate $\tau_q^{-1}$ and diffuses with
the effective diffusion constant
$D_{eff}^{\rho\rho}=2 D \Sgi d^2/\pi^2\tau_q = 4\lambda_D^2/\tau_q$.
The deviation from the equilibrium conductivity relaxes with the rate
$2 r/\tau_q=1/\tau_{rec}$, diffuses with the effective constant
$D_{eff}^{\sigma\sigma}= D \Sgii d^2/\pi^2 \tau_q
= 2\Sgii\lambda_D^2/\tau_q$ and is advected with the effective velocity
$v_{\sigma} = \alpha \Dg E_0 d/\pi\tau_q$. Both quantities are
coupled by further relaxational, advective and diffusive terms.
For equal mobilities ($\gamma=1, \Dg=0$) only the diagonal relaxation
and diffusion constants and the offdiagonal advective terms survive.
The nonlinear terms $\propto \dsigo\drhoo$ and $\propto \dsigo^2$ in
\refkl{nontriv_sig}
and \refkl{nontriv_rho} increase the respective relaxations by factors
$1+\dsigo$ and $1+\dsigo/2$. The term $\propto \drhoo^2$ contains
the Coulomb repulsion \cite{perez} and a part of the nonlinear recombination.
Equations \refkl{scal_bc} give the BCs for $\drhoo, \ \sigma_0=1+\dsigo$ and
$E_0$.

Clearly, the trivial basic state is incompatible with the
blocking BCs
\refkl{scal_bc} because the BCs are not homogeneous in $\drhoo$ and $\dsigo$.
For 'ohmic' BCs
$J_z=\sigma_{eq}E_z$ and $J_z^+-J_z^-=e(\mu_{\perp}^+-\mu_{\perp}^-)E_z$
the BCs \refkl{scal_bc} are homogeneous with $\sigma_0$ replaced by
$\delta\sigma_0$ and the trivial basic state is a solution.
For the blocking or other general BCs boundary layers
of accumulated charge appear, which influence the impedance of the system.
These effects will be discussed in future work. Here we assume that their
influence on the stability of the basic state is small (see below).

The linearised equations of the electrodiffusion model are obtained by
inserting
the decomposition \refkl{decomposition} into the scaled equations
\refkl{scal_rho} and \refkl{scal_sig}. The two electrical equations of the
resulting
set of equations for $\tilderho$ and $\tildesig$ are
(primes on the basic-state quantities denote $z$ derivatives and $\opsig$ is
defined in analogy to $\opeps$)
\clearpage
\begin{eqnarray}
\label{lin_rho}
P_1\d{t}(\opeps \phi    &+&    \epsa E_0\d{x}n_z)                    \nonumber
\\
  &=& \left[- \sigma_0 \opsig + \sigma'_0\d{z}
            - 2 D \Sgi \opsig \opeps \right]\phi
  -  \left[E_0\d{z} + \drhoo + \alpha^{-1} D \Dg \opsig \right]\tildesig
\\
  &+& \left\{ -\siga\sigo E_0 + D \left[\alpha^{-1}\Dg\siga\sigma_0'
                         + 2\Sgi(\siga\drhoo' -
                                  \opsig\epsa E_0)\right]\right\}\d{x}n_z
  - P_1\drhoo' v_z, 						     \nonumber \\
\label{lin_sig}
P_1\d{t}\tildesig =
  & & \alpha\left\{ -\alpha \Sgi\left[
       (E_0\opeps-\drhoo')\d{z} +
        \drhoo(\opsig + \opeps + 2 r \opeps)
                              \right] \right.
\nonumber \\
  &-& \left.  \Dg\left[
            \sigma_0\opsig-\sigma_0'\d{z}+D\Sgi\opsig\opeps-r\sigma_0
            \opeps \right]
      \right\} \phi
\nonumber \\
  &-& \left[\alpha\Dg (E_0\d{z}+\drhoo(1-r))
            + D\Sgii \opsig + 2 r \sigma_0 \right] \tildesig         \nonumber
\\
  &+& \left\{ \alpha\Dg \left[
        -\sigo\siga E_0 + D \Sgi (\siga\drhoo'-\opsig\epsa E_0) + r\sigo\epsa
E_0
                       \right] \right.
\\
  &+& \left. \alpha^2\Sgi E_0 \left[
        -\drhoo\siga - \epsa (2\drhoo(1+r) + E_0\d{z})
                      \right]
        + D \Sgii\siga\sigma_0' \right\} \d{x}n_z
        - P_1\sigma_0' v_z, 					    \nonumber
\end{eqnarray}
with the (homogeneous) BCs
\begin{eqnarray}
\label{lin_bc_i}
[E_0-\alpha^{-1}D\Dg\d{z}]\tildesig + [\sigma_0-2 D
\Sgi\d{z}^2]\tilde{E}_z&=&0, \\
\label{lin_bc_ii}
[D\d{z}-\alpha E_0\Dg]\tildesig
   - \alpha [\Dg\sigma_0 + 2\alpha \Sgi (\drhoo + E_0\d{z})]\tilde{E}_z
                                        &=& 0.
\end{eqnarray}
Since for director modes $\d{z}\approx \pi/d$ and $D<<1$ the diffusive parts
can be
neglected leading to $\tildesig/\tilde{E}_z=-\sigma_0/E_0$ and
$(\drhoo + E_0\d{z})\tilde{E}_z=0$,
which can be fulfilled by the BCs of the standard
model $\d{z}\tilde{E}_z=\tilderho = 0$ for $\drhoo=0$.
The linearized director and momentum-balance equations of the standard model
are changed by some terms from the nontrivial basic state. They are given
(other scaling) in Ref. \cite{kramer_jphys}, Eqs. (3.2) with $V(t)$ replaced
by $V(t)+\pi\dEo(z,t)$ and the volume-force term $\propto q\phi$ in (3.2e)
($iq$ corresponds to $\d{x}$) supplemented by a contribution
$\drhoo'\epso q \phi$.

For $\alpha\to 0,\ D\to 0$ and without the nontrivial parts
of the basic state one has $\tildesig \to 0$ and Eq. \refkl{lin_rho} reduces to
$(P_1\d{t}\opeps + \opsig)\phi = - \pi^{-1}(P_1\epsa\d{t} +
\siga)V'(t)\d{x}n_z$,
the continuity equation (3.2a) in \cite{kramer_jphys}.

The BCs for $\drhoo, \ \sigma_0=1+\dsigo$ and
$E_0$ are the same as \refkl{scal_bc}.
Neglecting the small field deviations $\dEo$ which are of the order of
$\lambda_D/d$, $E_0 \approx \pi^{-1}V'(t)$ is spatially constant and
the first two terms on the rhs. of \refkl{nontriv_sig} and \refkl{nontriv_rho}
are linear in the deviations from the standard-model-basic state.
\vspace{8mm}

\sectionILCC{4. ANALYTICAL RESULTS IN A SIMPLE CASE}
In this section we show that the linear equations
\refkl{lin_rho} - \refkl{lin_bc_ii} can lead to a Hopf bifurcation.
We assume blocking BCs and boundary layers with a thickness much
smaller than the cell thickness and in the rest of the volume
an essentially trivial basic state, i.e. we adopt
$\sigma_0=1, \pi E_0(t) = V'(t) = \sqrt{2 R}\cos\omega t$ and $\drhoo=0$.
Further we neglect diffusion and restrict ourselves to the normal-roll
regime by setting
$(\phi\vonrt , \tildesig\vonrt, n_z\vonrt) =
e^{iqx}(\barphi(z,t), \barsig(z,t), \barn_z(z,t))$ and
$n_y=v_y=0$.
The fluid velocities $v_x, v_z$ in the director equation for $n_z$
are eliminated adiabatically. This is
justified because the viscous
relaxation time $\tau_{visc} = \rho_m d^2/(\alpha_4/2) \approx 10^{-5}$s
is usually much shorter than the other relevant time scales
\cite{pesch,treiber_fluct}.
With these assumptions one obtains from \refkl{lin_rho}, \refkl{lin_sig} and
the director and fluid equations from the standard model
\begin{eqnarray}
\label{lintriv_rho}
\left[ P_1\d{t}\opepsq + \opsigq \right] \barphi
   + E_0(t)\d{z} \barsig
   + \left[ P_1\epsa(\dot{E}_0(t) + E_0(t)\d{t}) + \siga E_0(t) \right]
iq\barn_z
             &=& 0, \\
\label{lintriv_sig}
\alpha \left[ \alpha E_0(t)\Sgi \opepsq\d{z} + \Dg (\opsigq - r\opepsq)\right]
                         \barphi
   + \left[ P_1\d{t} + \alpha \Dg E_0(t)\d{z} + 2 r \right] \barsig
                     & & \nonumber \\
  \hspace{5mm} + \alpha E_0(t) \left[ \Dg (\siga - r \epsa) + \alpha \Sgi\epsa
E_0(t)\d{z}\right]
                           i q \barn_z
             &=& 0, \\
\label{lintriv_n}
\frac{\hat{L}_{n\phi}}{\siga}\pi^2E_0(t) \barphi
    + \left[\d{t} - \hat{\lambda}_0 - \frac{\epsa}{\siga}\pi^2E_0^2(t)\right]
                           i q \barn_z
             &=& 0,
\end{eqnarray}
where $\opepsq=-\d{z}^2 + (1+\epsa)q^2$, \
$\opsigq=-\d{z}^2 + (1+\siga)q^2$
(the operators $\opeps$ and $\opsig$
applied to the normal-roll state) and $\opKq = -\d{z}^2 + K_{33}q^2$.
The zero-field director relaxation operator $\hat{\lambda_0}$ is given by
\cite{treiber_fluct}
\begin{equation}
\label{lambdao}
\hat{\lambda_0} = -\frac{\opKq}{K_{eff}}(1-\hat{L}_{nn})^{-1},
\end{equation}
For simplicity the $z$-dependence of the velocity $v_z$ is expressed already
here in terms
of a normalized Galerkin mode $f_v(z)$ which gives
\begin{eqnarray}
\label{Lnphi}
\hat{L}_{n\phi} & = &\frac{\opepsq-\epsa L_{nn}q^2}{1-L_{nn}}, \\
\label{Lnn}
L_{nn} &=&  q^4 \left\{ \int^{\pi/2}_{-\pi/2}dz \ f_v^{*}(z)\left[
   \frac{\alpha_4 + \alpha_6}{2}(q^2-\d{z}^2)^2 + q^2(q^2-(1+\alpha_1)\d{z}^2)
   \right] f_v(z) \right\}^{-1}
\end{eqnarray}
with the anisotropic shear
viscosities $\alpha_i, i=1 ... 6$  of the NLC. Some small terms $\propto
\alpha_3$
have been omitted.
The BCs \refkl{lin_bc_i}, \refkl{lin_bc_ii} without diffusion
can be combined to give
\begin{equation}
\label{lintriv_bc}
   \d{z}^2\barphi = 0, \ \ \d{z}\barphi = E_0(t) \barsig.
\end{equation}
The first BC is consistent with the ansatz for the field
inhomogeneity of the symmetric-conductive mode of the standard model
(Type I A of \cite{kramer_jphys,zimmermann}).
Since then $\d{z}\barphi \neq 0$, the second BC implies
the conductivity perturbation to be nonzero at the
boundaries, i.e the lowest-order
mode of the conductivity is antisymmetric in $z$.

To get analytical results we use the lowest-order Galerkin ansatz for
the symmetric-conductive director mode,
$i q \barn_z= \nhocho\cos kz, \barphi=(\phihochi e^{i\omega t} +
  \phihochmi e^{-i\omega t}) \cos kz$,
the antisymmetric mode for the conductivity
$\barsig = \sighocho \sin kz$ to which the $\phi$ equation
\refkl{lintriv_rho} couples $\nhocho$ and $\phi^{(\pm1)}$ via the
$E_0\d{z}$ term, and the first Chandrasekhar function \cite{kramer_jphys} for
$f_v(z)$.
The $\sigma$ equation \refkl{lintriv_sig} couples at lowest order also to
symmetric conductivity modes $\propto e^{\pm i \omega t}\cos kz$
but they do not satisfy the second BC \refkl{lintriv_bc}.
Inserting this ansatz into \refkl{lintriv_rho}-\refkl{lintriv_n}
one obtains four coupled
equations for $\nhocho, \phihochmi, \phihochi$ and $\sighocho$.
Assuming for the charge relaxation rate $\tau_q^{-1} >> \tau_{rec}^{-1},
\tau_d^{-1}$ one can eliminate adiabatically the induced potentials
$\phi^{(\pm 1)}$ and obtains a $2\times 2$ system for the
symmetric director/antisymmetric conductivity modes,
\begin{equation}
\label{lingalerkin}
\begin{array}{rrr}
\left[ \d{t}-\lambda_{\sigma}(R,\omega)\right] \sighocho
   & + \left[ \frac{
       R \tilde{\alpha}^2(\siga\frac{\epsilon_q}{\sigma_q}-\epsa)}{
          1+\omega'^2} \right] \nhocho
   & = 0, \\
\left[ -\frac{L_{n\phi} R}{
          (1+\omega'^2)\sigma_q\siga} \right] \sighocho
    & + \left[\d{t}-\lambda_n(R,\omega)\right] \nhocho
    & = 0.
\end{array}
\end{equation}
The numbers $\epsilon_q, \sigma_q, K_q$ and $\lambda_0$ are the corresponding
operators with $\d{z}^2\to -1$. The normalized external frequency is
$\omega'=\omega_{ext}\tau_q\epsilon_q/
\sigma_q$, where $\epsilon_q/\sigma_q$varies  typically from
0.6 to 0.9.
The relaxation rates $\lambda_{\sigma}$ and $\lambda_n$ of the
uncoupled $\sigma$ mode and of the conductive director
mode of the standard model are given by
\begin{eqnarray}
\label{lambdasig}
\lambda_{\sigma} &=& -\left(\tilde{r} + \frac{R\tilde{\alpha}^2\epsilon_q}{
          (1+\omega'^2)\sigma_q} \right), \\
\label{lambdan}
\lambda_n &=& \lambda_0 + R \left(
   \frac{L_{n\phi}}{(1+\omega'^2)\sigq} + \frac{\epsa}{\siga}\left(
      1+\frac{\omega'^2 L_{n\phi}}{(1+\omega'^2)\epsq} \right)
               \right)
   := \lambda_0 \left(1 - \frac{R}{R_{c0}}\right),
\end{eqnarray}
where $R_{c0}$ is the threshold of the standard model.
Besides the two control parameters $R$ and $\omega'$,
Eq. \refkl{lingalerkin} contains two parameters describing the migration and
recombination effects,
\begin{eqnarray}
\label{tildealpha}
\tilde{\alpha}^2 &=& \frac{\alpha^2\Sgi}{P_1\pi^2} =
  \frac{\mu_{\perp}^{+}\mu_{\perp}^-\gamma_1\pi^2}{\sigma_a d^2}, \\
\label{tilder}
\tilde{r} &=& \frac{2r}{P_1} = \frac{\taudo}{\tau_{rec}}.
\end{eqnarray}
The coupling of the director to the conductive mode is proportional to the
square of an effective mobility parameter $\tilde{\alpha}$
which is $\alpha$ with $\mu$ and $\tau_q$ replaced by the geometrical means
of $\mu_{\perp}^{\pm}$ and $\tau_q, \tau_d$, respectively.
Experimentally $\tilde{\alpha}$ can be influenced by varying
$\sigp d^2$, see Table III.

The growth rate of modes $\propto e^{\lambda t}$ in \refkl{lingalerkin}
is given by
\begin{eqnarray}
\label{lambda}
\lambda &=&  \frac{\lambda_{\sigma}+\lambda_n}{2} \pm \sqrt{\frac{
  (\lambda_{\sigma}-\lambda_n)^2}{4} - C^2\left(\frac{R\tilde{\alpha}}{
      1+\omega'^2}\right)^2} \\
\label{C}
C^2 &=& \frac{\epsq^2\left(1-\frac{\epsa}{\epsq}\frac{\sigq}{\siga}\right)
                 \left(1-\frac{\epsa}{\epsq}L_{nn}q^2 \right)}{
            \sigq^2(1-L_{nn})},
\end{eqnarray}
where $C$ is (apart from a weak $q$ dependence via $q_c(\omega')$) a
fixed constant for a given NLC. For MBBA
$C^2(q=q_c)=1.19$ with $q_c=1.51$ for $\omega' << 1$
and $C^2(q=q_c)=2.71$ with
$q_c\to \infty$ for $\omega' \to \omega'_{cutoff}$. Setting
$\epsa = 0$ to simulate the material I52, see \cite{dennin}, one has
$C^2=0.91, q_c=1.21$, independent of frequency (in this approximation).

A Hopf bifurcation is characterized by a nonzero imaginary part of
$\lambda$ at threshold, Re$\lambda(R_c,q_c)=0$, Im$\lambda(R_c,q_c)
  = \omega_H \neq 0$ with
\begin{equation}
\label{hopf_frequency}
\omega_H = 2\pi f_H =
\frac{R_c\tilde{\alpha} C}{1+\omega'^2}\sqrt{1-\frac{1}{C^2}
\left( \frac{\tr}{\talpha}\frac{1+\omega'^2}{R_c} +
       \talpha \frac{\epsilon_{q_c}}{\sigma_{q_c}}\right)^2}.
\end{equation}
The Hopf condition, namely, that the root be positive,
leads essentially to conditions for $\tr/\talpha$ and $\talpha$.
For typical
material and system parameters (see Tables II and III), one usually has
$\talpha \epsq/(\sigq C) << 1$ so
we are left with an upper limit for the recombination
rate,
\begin{equation}
\label{hopf_condition}
\tr < \frac{\talpha R_c C}{1+\omega'^2}.
\end{equation}
Since $\tr/\talpha \propto (\sigp d^2)^{3/2}$ (Table III) this condition
gets weaker for thinner and cleaner cells; for materials with $\epsa<0$ it is
always
fulfilled near the cutoff frequency $\omega'_{cutoff}$, where
$R_c\to \infty$.
In the special case $\epsa \to 0$, relevant for experiments with
the nematic material I52 \cite{dennin}, Eq. \refkl{hopf_frequency}
simplifies to
\begin{equation}
\label{hopf_simple}
2\pi f_H(\epsa=0) = -\lambda_0\talpha\sqrt{(1-L_{nn}) -
  \left(\frac{-\tr}{\lambda_0\talpha} + (1-L_{nn})\talpha\right)^2}
\end{equation}
If the Hopf conditions are well satisfied, the Hopf frequency
is essentially $2\pi f_H = -\lambda_0\talpha(1-L_{nn})^{1/2}$ proportional to
$\sigp^{-1/2}d^3$. For MBBA parameters with
$\epsa$ changed to zero (I52) this is in physical units
$f_H\tau_d \approx 0.4\talpha$.
\vspace{8mm}

\begin{figure}
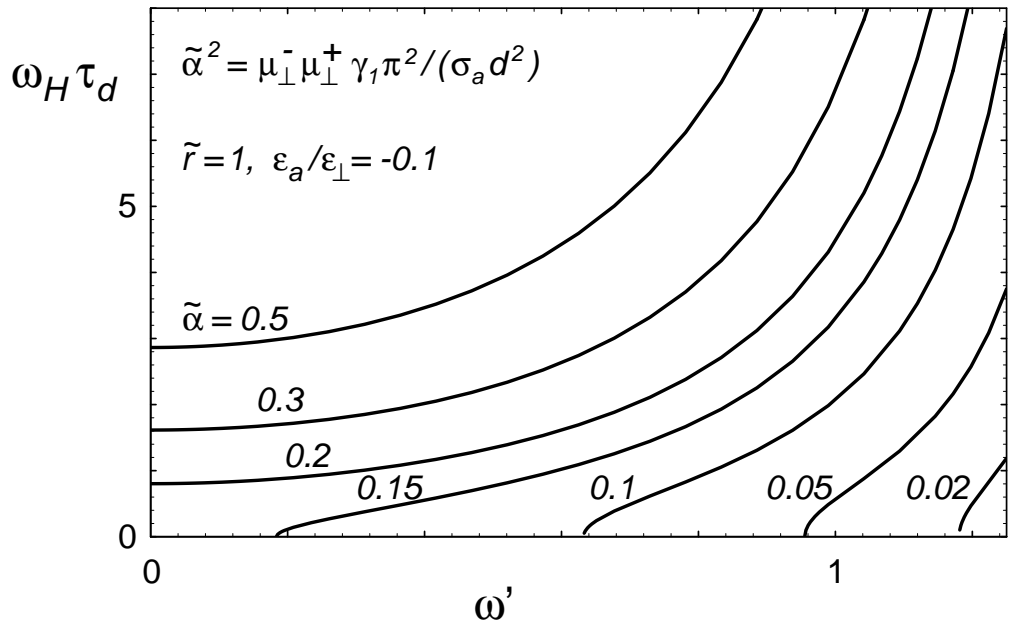

\caption{Hopf frequency $\omega_H = 2\pi f_H$, normalized to the director
relaxation time, as function of the normalized external frequency
$\omega'=\omega_{ext}\tau_q \epsq/\sigq$
for the analytical model of Sec.4 with MBBA I parameters and
$\tilde{r}=\tau_d/\tau_{rec}=1$.
$\tilde{\alpha}$ is the mobility parameter. The cutoff frequency corresponds
to $\omega'=1.4$.
}
\end{figure}

\begin{figure}
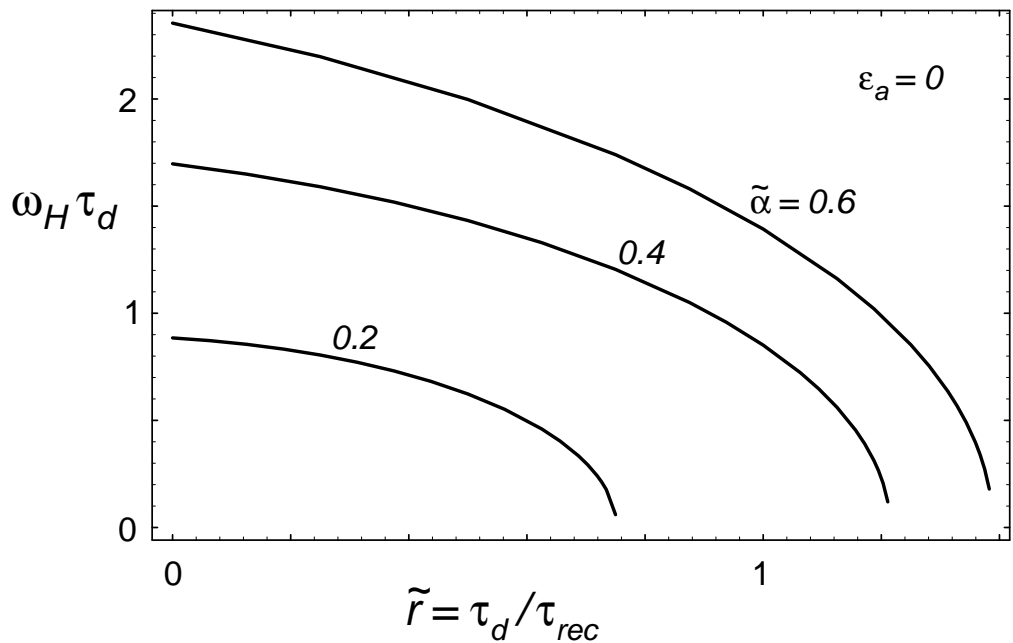

\caption{Hopf frequency as function of the recombination
rate for MBBA I parameters with $\epsa$ set to zero (to simulate $I52$)
and different values of $\tilde{\alpha}$.
}
\end{figure}
\sectionILCC{5. DISCUSSION}
We presented a new theory for electroconvection by replacing the ohmic
conductivity
of the standard model with drift and diffusion of two species
of charge carriers. The conductivity itself becomes a new
dynamically active variable, if the recombination time is not  small
compared to the charge or the director relaxation
time. We demonstrated by a simple analytical calculation
that this new degree of freedom leads generically
to a Hopf bifurcation, if the ratio $\tilde{\alpha}/\tilde{r}$ of the
mobility parameter $\tilde{\alpha}$ and the recombination
parameter $\tilde{r}$ is large enough.

Since the ratio $\tilde{\alpha}/\tilde{r}$ is proportional to
$(\sigp d^2)^{-3/2}$, the above Hopf condition is fulfilled for thin and
pure (small conductivity $\sigp$) samples, in qualitative
agreement with experiments on MBBA, where Hopf is only found for
thin \cite{rehberg_travV,rehberg_travMBBA} and clean cells \cite{buka},
and with experiments on I52 \cite{dennin} where the conductivity could be
changed
and Hopf was found only for low conductivities.
For materials with $\epsa < 0$ (e.g. MBBA) the Hopf condition gets weaker (and,
if it is fullfilled, the Hopf frequency increases) with increasing external
frequencies $\omega_{ext}$. For $\epsa=0$ the Hopf
condition is nearly independent of $\omega_{ext}$. In fact in I52, Hopf
is observed in sufficiently clean samples for all frequencies \cite{dennin}.

In Fig.1 our results for the (normalized) Hopf frequency as a function of
the (normalized) external frequency are plotted for MBBA for various values of
$\tilde{\alpha}$ and $\tilde{r}=1$. The behavior found in the experiments of
ref.\cite{rehberg_travV} is similar to our results for
$\tilde{\alpha}\approx 0.1$ (at $\tilde{r}=1$), whereas the behavior found in
ref.\cite{rehberg_travMBBA} is quite similar to our results for
$\tilde{\alpha}\approx 0.17$. Figure 2 shows the Hopf frequency as a
function of $\tilde{r}$ for MBBA with $\epsa = 0$ for three values of
$\tilde{\alpha}$.

\vspace{12mm}

\vspace{3mm}
\noindent
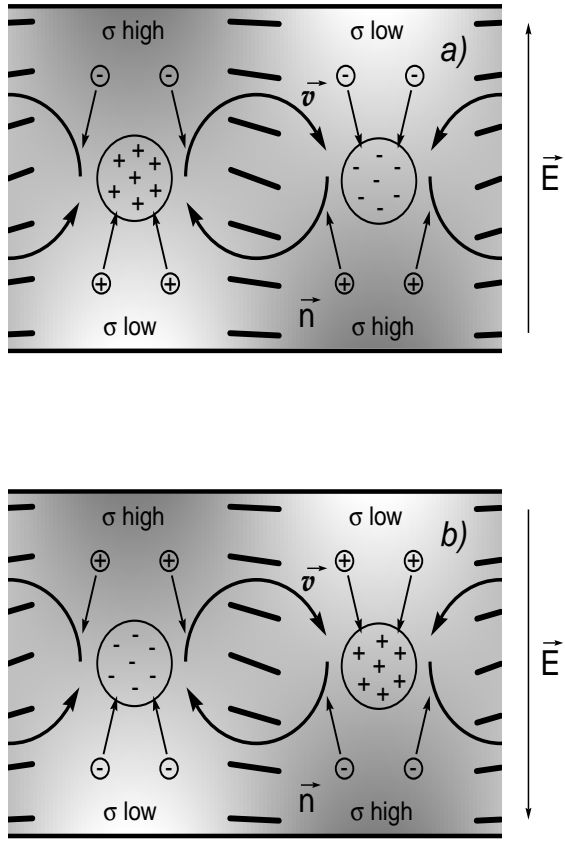
Figure 3: Sketch of the spatial distribution of the physical quantities
inside the nematic layer (conductive regime). The straight arrows indicate
the contributions to the current carried by each species. The shading
illustrats the conductivity mode (dark $=\sigma$ high, light $=\sigma$ low)
\clearpage

A qualitative understanding of the mechanism leading to the Hopf bifurcation
can be obtained by noting that the conductivity mode describing rearrangement
of
the total charge carrier density (each species weighed by its mobility)
on the time scale of $\tau_{mig}$ reduces the
charge density produced by the standard Carr-Helfrich charge focussing
mechanism.
This antagonistic cross coupling of $\sigma$ and $\rho$  can already be
identified in Eqs.\refkl{bipol_rho},\refkl{bipol_sig} or \refkl{scal_rho},
\refkl{scal_sig} (one may choose e.g. equal mobilities, $\gamma = 1$, so
that for $D=0, r=0$ one is left only with cross-coupling terms), and carries
through to Eqs.\refkl{lingalerkin}, where it is easily seen to lead
to an increase of the
threshold (note that in Eqs.\refkl{lingalerkin} all effects from
the standard theory are included in the relaxation rate $\lambda_n$). Figure 3
gives a sketch of the resulting distribution of the physical quantities
inside the layer at two instances separated by half of an external period. If
the
time scale of the cross coupling effect is sufficiently long, the system
acts like a spring and destabilization occurs in an oscillatory manner.
The linear modes of the system then describe left and right traveling rolls
(or "waves"), although in principle one could also have the more complicated
superposition to standing waves.

Since we made crude approximations (adiabatic elimination of the charge,
trivial basic state, lowest-order Galerkin expansion, normal-roll state
also for I52) we can expect
the above predictions to agree only qualitatively with the experiments.
Calculations with the full basic state and
linear equations are planned in the future. Then it should be possible,
to determine the recombination rate constant and the mobility ratio of the two
species by a fit of the predicted Hopf frequency to the measured one.
To get information about the basic state one can compare
the predicted cell impedance with measurements. A weakly nonlinear analysis
of the new model should determine the parameter ranges where
the predicted Hopf bifurcation is continuous or hysteretic. Finally one
should be able to compute all parameters of the universal Ginzburg-Landau
description valid near threshold.

We wish to thank A. Hertrich and W. Pesch for helpful discussions, M. Dennin
for making available his experimental results prior to publication,
and A. Buka for a critical reading of the manuscript. One of us (M.T.)
is grateful to the Arizona Center for Mathematical Sciences at Tucson,
where part of this work was performed, for its hospitality. Financial support
by Deutsche Forschungsgemeinschaft
(SFB 213, Bayreuth) and Stiftung Volkswagenwerk is gratefully acknowledged.

\renewcommand{\baselinestretch}{1} \small\normalsize


\begin{thebibliography}{99}

\bibitem{kramer_jphys} E.\ Bodenschatz, W.\ Zimmermann and
L.\ Kramer, J. Phys. France {\bf 49}, 1875 (1988);
L.\ Kramer, E.\ Bodenschatz, W. Pesch, W. Thom and W. Zimmermann, Liquid Cryst.
{\bf 5}, 699 (1989).
%
\bibitem{zimmermann}
W. Zimmermann in {\sl Nematics: Mathematical and Physical Aspects},
J.-M. Coron, J.M. Ghidaglia, and F. Helein, eds., NATO ASI
Series C - Vol.332 (Kluwer Acad. Publishers, Dordrecht, 1991), pp.401.
%
\bibitem{rehberg_fk}
I. Rehberg, B.L. Winkler,
 M. de la Torre Juarez, S. Rasenat, and W. Sch{\"o}pf,
 {\sl Festk{\"o}rperprobleme-Advances in Solid-State physics}
 {\bf 29}, 35 (1989).
%
\bibitem{pesch} L. Kramer and W. Pesch, {\sl Convective Instabilities
in Nematic Liquid Crystals}, to appear in {\sl Annual Review of Fluid
Mechanics}{\bf 27}, 1995.
%
\bibitem{kai} S. Kai and K. Hirakawa, Prog. Theor. Phys. Suppl.
{\bf 64}, 212 (1978).
%
\bibitem{rehberg_travV} I. Rehberg, S. Rasenat, and V. Steinberg,
Phys. Rev. Lett {\bf 62}, 756 (1989).
%
\bibitem{joets} A. Joets and R. Ribotta, Phys. Rev. Lett. {\bf 60} 2164 (1988).
%
\bibitem{rehberg_travMBBA} I. Rehberg, S. Rasenat, J.Fineberg, M. de la Torre
and V. Steinberg, Phys. Rev. Lett {\bf 61}, 2449 (1988).
%
\bibitem{dennin} M. Dennin, D.S. Cannell, and G. Ahlers,
this proceedings.
%
\bibitem{rehberg_fluctPRL} I. Rehberg, S. Rasenat, M. de la Torre Juarez,
  W. Sch{\"o}pf, F. H{\"o}rner, G. Ahlers, and H.R. Brand,
        Phys. Rev. Lett {\bf 67}, 596 (1991).
%
\bibitem{erickson} Ericksen, J. L., Arch. Ration. Mech. Analysis {\bf 23},
   266 (1966).
%
\bibitem{leslie}  F. M. Leslie, Quart. J. Mech. Appl. Math. {\bf 19},
   357 (1966).
%
\bibitem{richardson}  R. Chang and J.M. Richardson,
  Mol. Cryst. Liq. Cryst {\bf 28}, 189 (1973).
%
\bibitem{turnbull}
  R.J. Turnbull, J. Phys. {\bf D6}, 1745 (1973).
%
\bibitem{naito_1}  H. Naito, K. Yoshida, and M. Okuda,
  J. Appl. Phys {\bf 73}, 1119 (1993).
%
\bibitem{naito_2}  H. Naito, M. Okuda, and A. Sugimura
  Phys. Rev. {\bf A44}, 3434 (1991).
%
\bibitem{naito_3} A.Sugimura et al., Phys. Rev. {\bf B43}, 8272 (1991).
%
\bibitem{naito_buda} S. Murakami, H. Naito, M. Okuda, and A. Sugimara,
  Presented (Poster K-P23) at the 15th ILCC in Budapest, 1994.
%
\bibitem{felici}
  N. Felici, Revue generale de l' electricite {\bf 78}, 717 (1969).
%
\bibitem{treiber_fluct}
M. Treiber and L. Kramer,
   Phys. Rev. {\bf E94}, 3184 (1994).
%
\bibitem{sigasource}
  D. Diguet et al., Compt. Rend. {\bf 271B}, 954 (1970)
%
\bibitem{perez} A.T. Perez and A. Castellanos, Phys. Rev. {\bf A40}, 5844
(1989).
%
\bibitem{buka} H. Richter, A. Buka and I. Rehberg, "Convection in a
Homeotropically Aligned Nematic", to be published (homeotropically aligned
specimens with $\epsa<0$ behave in many respects similar to planarly aligned
ones).
\end{thebibliography}
\end{document}